\documentclass[aip,apl,preprint]{revtex4-1}

\usepackage{graphicx}
\usepackage{amsmath}
\usepackage{amssymb}
\usepackage{hyperref}
\usepackage[version=3]{mhchem}

\begin{document}

\title{Atomistic origin of glass-like \ce{Zn4Sb3} thermal conductivity}

\author{Xudong Li}
\affiliation{Frontier Institute of Science and Technology, Xi'an Jiaotong University, 710054, Xi'an, China}
\affiliation{State Key Laboratory for Mechanical Behavior of Materials, Xi'an Jiaotong University, 710049, Xi'an, China}

\author{Jes\'{u}s Carrete}
\affiliation{CEA-Grenoble, 17 Rue des Martyrs, Grenoble 38000, France}

\author{Jianping Lin}
\affiliation{State Key Laboratory for Mechanical Behavior of Materials, Xi'an Jiaotong University, 710049, Xi'an, China}

\author{Guanjun Qiao}
\affiliation{State Key Laboratory for Mechanical Behavior of Materials, Xi'an Jiaotong University, 710049, Xi'an, China}

\author{Zhao Wang}
\email{wzzhao@yahoo.fr}
\affiliation{Frontier Institute of Science and Technology, Xi'an Jiaotong University, 710054, Xi'an, China}

\begin{abstract}
Zinc antimony stands out among thermoelectrics because of its very low lattice  thermal conductivity, close to the amorphous limit. Understanding the physical reason behind such an unusual crystal property is of fundamental interest for the design of new thermoelectric materials. In this work we report the results of atomistic computer simulations on experimentally determined  $\beta-$\ce{Zn4Sb3} structures. We find a remarkably anharmonic behavior of Zn atoms that could be responsible for the low thermal conductivity of \ce{Zn4Sb3}: their movement, better explained as diffusive, does not contribute to thermal conduction. Moreover, phonon transport is impeded by a lack of coupling between Zn and Sb atoms in crystalline positions.
\end{abstract}

\maketitle

As one of the best thermoelectric materials at moderate temperature,\cite{Snyder2004} \ce{Zn4Sb3}'s remarkable performance (as gauged by its high dimensionless thermoelectric figure of merit, $ZT$) is mainly derived from its glass-like lattice thermal conductivity, below $1.0\,\mathrm{W\,m^{-1}\,K^{-1}}$. Such a low thermal conductivity approaches the theoretical lower bound known as the amorphous limit, introduced by Slack \cite{slack} and refined by Cahill and coworkers.\cite{cahill_lower_1992}. The idea behind this theoretically minimal thermal conductivity is a physically plausible lower bound to phonon mean free paths, \cite{wang2011absence} such as half a wavelength. Crystalline compounds with thermal conductivities close to \cite{morelli_intrinsically_2008} or even below \cite{chiritescu_ultralow_2007} this theoretical limit are few and usually display interesting physics.

A straight path to the design of efficient thermoelectrics may lie in the extremely poor thermal conduction of \ce{Zn4Sb3}. Thus, many attempts at understanding this feature have been made by placing \ce{Zn4Sb3} under microscopes.\cite{Gault2010,Rauwel2011,Toberer2010,Prytz2009} For instance, using single-crystal X-ray and powder-synchrotron-radiation diffraction, Snyder and coworkers have found that the hexagonal unit cell of \ce{Zn4Sb3} containing at least three interstitial Zn atoms follows the rules of valence compounds, and that Zn occupancy at the strongly bonded crystal site reaches only about $90\%$.\cite{Snyder2004} This has provided significant insight for explaining the unusual chemical and physical properties of this material. Specifically, interstitial atoms are suspected to be an extremely effective mechanism for reducing thermal conductivity by introducing disorder.\cite{Schweika2007,Chen2011}

Despite many attempts, the precise mechanism behind the low lattice thermal conductivity of \ce{Zn4Sb3} is still a puzzle, due to the lack of atomic-scale information and the challenges posed by the study of dynamical effects in large systems through \textit{ab-initio} methods. For instance, scanning electron microscopy (SEM) and transmission electron microscopy (TEM) observations and density functional theory (DFT) calculations have provided evidence of Zn diffusion and nanovoid formation in $\beta-$\ce{Zn4Sb3},\cite{Rauwel2011a} but their connection to thermal conduction remains an elusive essential part. More generally, self-diffusion of ions in crystals has long been observed in experiments;\cite{Miller1942} its influence on conduction by phonons is also not well understood. Here we performed molecular dynamics (MD) simulations \cite{PhysRevB.83.125422} on pristine and interstitially doped $\beta-$\ce{Zn4Sb3} structures at different temperatures in order to obtain information at the atomic scale for understanding these effects. Our results qualitatively point to an intrinsic phonon anharmonicity induced by different dynamic behaviors of Sb and Zn atoms, which plays an essential role in enhancing the thermoelectric efficiency of \ce{Zn4Sb3}.

Recent experiments have found several different possible stoichiometries of $\beta-$\ce{Zn4Sb3}, and have shown that the lattice thermal conductivity is sensitive to these compositions.\cite{Toberer2010} In our experiments, two groups of samples were synthesized at two different starting Zn concentrations, $\approx 56.90$ at\% [nominally single-phase (SP) sample] and $\approx 55.53$ at\% [nominally zinc-poor (ZP) sample]. We measured the thermal diffusivity ($\alpha$), specific heat ($C_p$) and density ($\rho$) of the sintered samples for calculating the total thermal conductivity $\kappa$ using the relationship $\kappa = \rho \alpha C_p$. Electrical conductivities ($\sigma$) and Seebeck coefficients ($S$) were measured using a Linseis Seebeck Coefficient/Electric Resistance Measuring System (LSR-3). These data were used to estimate the electron contribution to the thermal conductivity ($\kappa_e$) through the Wiedemann-Franz relation $\kappa_e=L\sigma T$.\cite{May2009a,Johnsen2011} In keeping with recommendations from previous studies,\cite{Johnsen2011,Girard2011,May2009a} suitable values of the Lorenz number ($L$) were estimated from the reduced chemical potential $\eta$ obtained, in turn, from the Seebeck coefficient using the following expressions:

\begin{equation}
S=\frac{k_\mathrm{B}}{e} \left[\frac{2 F_1 \left(\eta\right)}{F_0 \left(\eta\right)}-\eta\right]
\label{eqn:S}
\end{equation}

\noindent and

\begin{equation}
L=\left(\frac{k_\mathrm{B}}{e}\right)^2 \frac{3 F_0\left(\eta\right)F_2\left(\eta\right)-4 F_1\left(\eta\right)^2}{F_0 \left(\eta\right)^2 },
\label{eqn:L}
\end{equation}

\noindent under assumptions of a single parabolic band and of dominance of electron scattering by acoustic phonons. Here $k_\text{B}$ is the Boltzmann Constant, $e$ is the elementary charge and $F_n \left(\eta\right) = \int _0^{+ \infty} x^n\left(e^{x-\eta}+1\right)^{-1}dx$ is the $n$th-order Fermi-Dirac integral. By way of example, using the above approach Girard and coworkers showed that the Lorenz number of Na-doped PbTe-PbS is reduced to nearly 65\% of the free-electron value $2.45 \times 10^{-8}\,\mathrm{\Omega\,W\,K^{-2}}$ at high temperatures.\cite{Girard2011} More generally, it is well known that $L$ can deviate significantly from such nominal value in contexts far removed from the average-scattering-time picture (degenerate semiconductors, very pure metals, systems with correlated electrons and so forth) and that its precise value depends on details of the density of states and the electron scattering mechanisms at play. Thus, it was considered safer to use the above $\eta$-dependent estimate. Finally, the lattice thermal conductivity  $\kappa_L$ was recovered by subtracting the electronic contribution from the total thermal conductivity, as $\kappa_L=\kappa-\kappa_e$.

\begin{figure}
\centerline{\includegraphics[width=11cm]{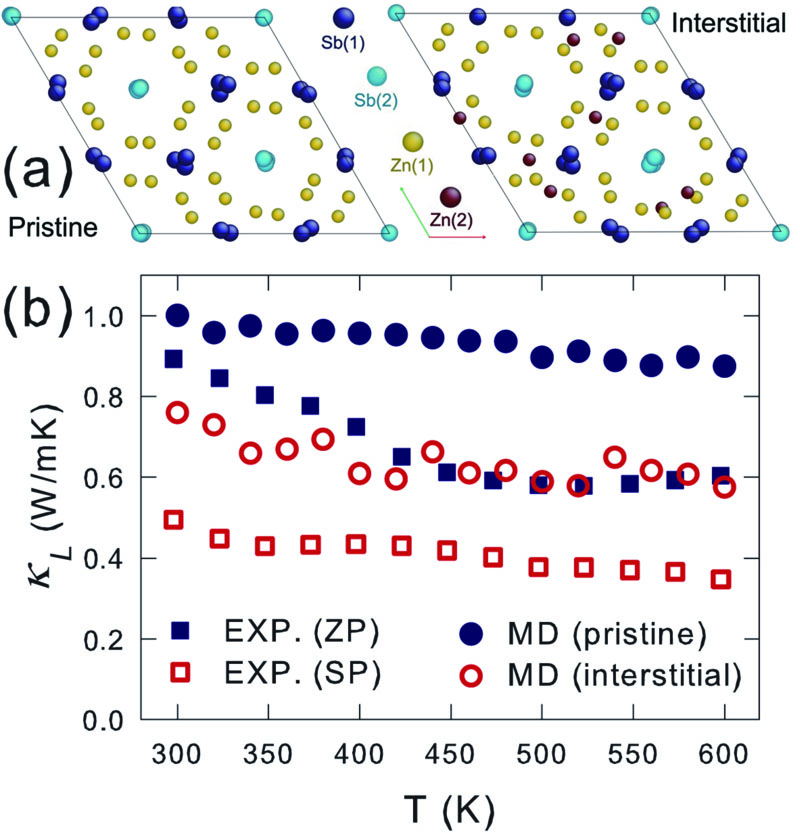}}
\caption{\label{fig1}
(a) \textit{left:} Pristine crystal cells of \ce{Zn4Sb3}, viewed from the $\left\langle 0001 \right\rangle$ direction. \textit{right:} A more stable configuration with interstitial Zn atoms. Circles in four different colors represent Zn at crystal sites [Zn(1)], Zn at glass-like sites [Zn(2)], and Sb at two different crystal sites [Sb(1) and Sb (2)], respectively. (b) Lattice thermal conductivity $\kappa_{L}$ as a function of temperature, experimental data plotted along with MD-computed values for samples in different stoichiometries.
}
\end{figure}

Taking into account the unavoidable precipitation of Zn due to the high diffusivity of this element, the Zn concentrations of our simulation samples were set to slightly lower values than those of the corresponding experimental samples. We started by building the complex hexagonal structure of $\beta$-\ce{Zn4Sb3} [Figure \ref{fig1} (a), left panel] in 3D space, with a concentration of Zn equal to $54.55$ at$\%$. Three Zn atoms were then randomly inserted into each unit cell at interstitial positions [Figure \ref{fig1} (a), right panel], in order to obtain the more stable interstitial structure of the SP sample. This yields a Zn concentration of $56.52$ at$\%$, that has been found to avoid phase separation.\cite{Rauwel2011} In our simulations we employed the  classical parallel molecular dynamics package LAMMPS.\cite{PLIMPTON1995} The simulation box comprised $12\times 12\times 12$ unit cells, with periodic boundary conditions applied in all three spatial directions. The equations of motion were integrated using the velocity Verlet algorithm with a time step of $0.5\,\mathrm{fs}$. A Nos\'e-Hoover thermostat was used to help the system reach thermal equilibrium at different temperatures before performing any statistical analysis. Atomic interactions were described by a pairwise potential that has been successfully applied to the study of the mechanical behavior of \ce{Zn4Sb3}.\cite{Li2011} We performed benchmark runs to check the quality of our potential function by simulating the characteristic $\beta$ to $\alpha$ phase transition around $250\,\mathrm{K}$, which manifests itself as a sharp step in the potential energy during a cooling process from the room-temperature $\beta$ phase.

We computed the lattice thermal conductivity of both the pristine and the interstitial structures using the Green-Kubo method.\cite{Hooverbook,schelling_comparison_2002} These values are plotted in Figure \ref{fig1} (b) along with experimentally measured data. Qualitatively, the same kind of decrease in $\kappa_{L}$ caused by interstitial atoms is observed in both experimental and simulation results when the concentration of Zn is increased. Note that the quantitative agreement between experiment and simulation is fairly good taking into account the $10-30\%$ uncertainty in $\kappa_L$ that can be expected from classical simulations.

\begin{figure}
\centerline{\includegraphics[width=11cm]{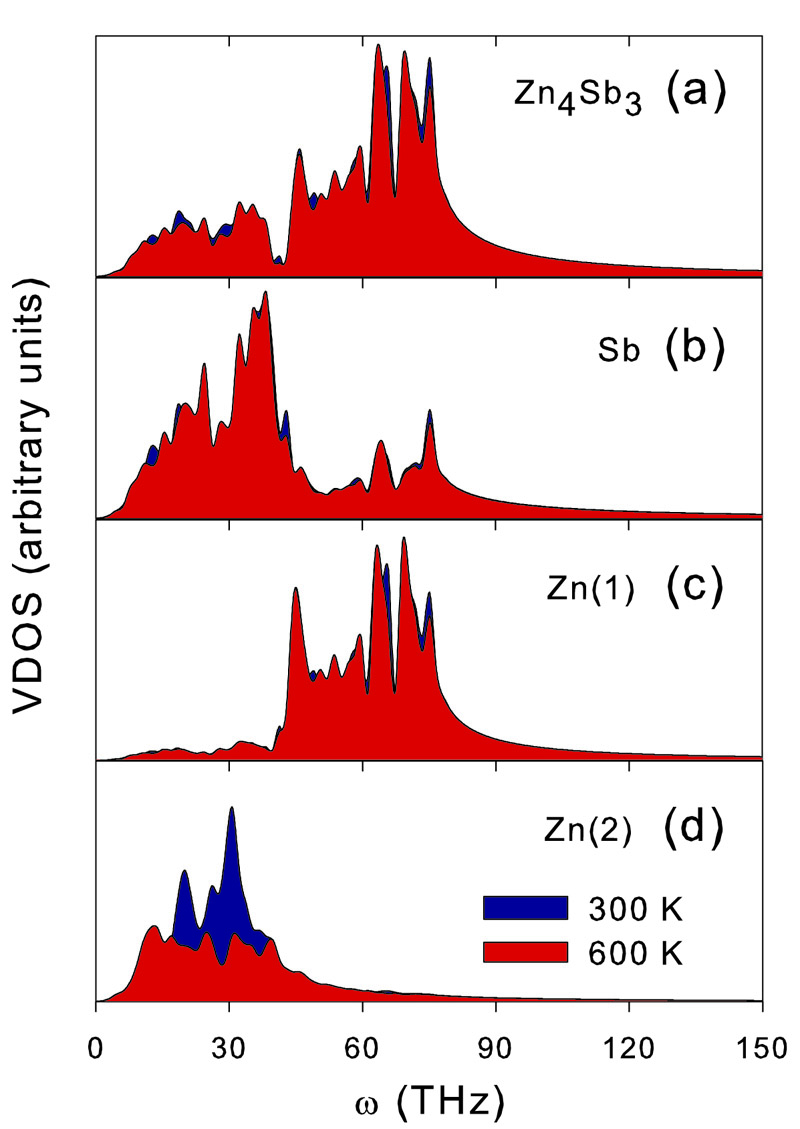}}
\caption{\label{fig2}
Vibrational densities of states (VDOS) for different atom types in the more stable interstitial $\beta$-\ce{Zn4Sb3} sample at two different temperatures.}
\end{figure}

We undertook more detailed analyses of the simulation results on the interstitial $\beta$-\ce{Zn4Sb3} sample, considered to be more stable at high temperature.\cite{Rauwel2011} We started by characterizing vibrational modes through the computation of the vibrational density of states (VDOS) projected on different atom sets. \cite{wang2010diameter} For an atom set $N$, the projected VDOS can be computed from the velocity auto-correlation function as

\begin{equation}
        \mathrm{VDOS}\left(\omega\right)=\left\vert \mathfrak{F}\left\lbrace
 \frac{\sum\limits_{i\in N} \mathbf{v}_i\left(t\right)\cdot\mathbf{v}_i\left(0\right) }{\sum\limits_{i\in N} \mathbf{v}_i\left(0\right)\cdot\mathbf{v}_i\left(0\right)};\omega\right\rbrace
\right \vert^ 2
\label{eqn:VDOS}
\end{equation}

\noindent where $\mathbf{v}_j\left(t\right)$ is the velocity of the $j$-th atom at time $t$, $\mathbf{v}_j\left(0\right)$ its initial velocity and $\mathfrak{F}$ denotes a Fourier transform. The data shown in Figure \ref{fig2} give insight into the efficient mechanism reducing thermal conduction in \ce{Zn4Sb3}. By comparing panels (a) to (d) it can be seen that Zn atoms at crystal sites contribute mainly to modes with frequencies in the range $50-100\,\mathrm{THz}$, whereas modes below $45\,\mathrm{THz}$ mainly involve Sb atoms, with a second-order contribution from Zn(2). This suggests that Zn(1) and Sb atoms are loosely coupled, which in itself may difficult heat conduction, but also significantly that the movement of Zn(2) atoms has very particular features that point to either rattling or diffusive motion. This impression is reinforced by the observation, in  Figure \ref{fig2} (d), of a very clear dependence on temperature of the vibrational density of states of Zn(2) atoms, \textit{i.e.} a high degree of anharmonicity. In fact, when $T$ is increased from $300\,\mathrm{K}$ to $600\,\mathrm{K}$ there is barely any change in panels (a) to (c), but the projected density of states in Figure \ref{fig2} (d) is smeared to such a degree that even peaks that are very marked at the lower temperature go unnoticed at $600\,\mathrm{K}$.  Not only is anharmonicity well known to be inversely correlated with thermal conductivity, \cite{lindsay_three-phonon_2008} but in this case it is so potent that it suggests that the movement of Zn(2) atoms is different from simple oscillation.

The influence of ionic diffusion on thermal conduction has long been a topic of interest in solid-state physics. Previous experiments have suggested that it could be a factor in reducing the thermal conductivity in \ce{Zn4Sb3}.\cite{Snyder2004,Toberer2010} A quantitative model for the temperature-dependent phonon spectrum of interstitial atoms in a system such as the one under consideration, built upon simplified assumptions, can be found in the work of Yamakage and Kuramoto,\cite{yamakage_temperature_2009} who considered a lattice formed by cage and guest ions. They showed that in an Einstein model the oscillation frequency of guest ions exhibits a $\propto T^2$ dependence at low temperatures and a less pronounced $\propto T^{1/4}$ behavior in the high-temperature limit. The fact that the peaks in the VDOSs presented in Figure \ref{fig2} do not experience such a shift, but instead show a much more dramatic loss of their harmonic structure, enables us to describe the behavior of this system as thermally-activated diffusion, in contrast to temperature-dependent oscillation.

In the Green-Kubo formalism used here, $\kappa_L$ is proportional to the integral

\begin{equation}
  \int\limits_0^{\infty} \left\langle J^{\left(z\right)}\left(t\right) J^{\left(z\right)}\left(0\right)\right\rangle dt.
\label{eqn:integral}
\end{equation}

\noindent The integrand is the auto-correlation function of the component of the heat current along the transport direction ($z$) at time $t$, $J^{\left(z\right)}\left(t\right)$. In a molecular-dynamics setting, this correlation is translated into a time average over discrete time steps of finite length $\Delta t$, that is,

\begin{equation}
\kappa\propto \sum\limits_{m=1}^M\frac{1}{L-m}\sum\limits_{n=1}^{L-m} J^{\left(z\right)}\left(n\Delta t\right) J^{\left(z\right)}\left(0\right)
\label{eqn:discrete}
\end{equation}

\noindent where $L$ is the total number of simulation steps and $M\Delta t$, with $M<L$, is a time cutoff chosen so as to approximate the improper integral in Eq. \eqref{eqn:integral}. The thermal current appearing in this formula is computed from forces and velocities,\cite{schelling_comparison_2002} and since it can be expressed as a sum over atoms, a fraction of the total current can be assigned to each component of the system, \textit{i.e.}, $J^{\left(z\right)}=J^{\left(z\right)}_{\mathrm{Zn\left(1\right)}}+J^{\left(z\right)}_{\mathrm{Zn\left(2\right)}}+J^{\left(z\right)}_{\mathrm{Sb}}$. In the same spirit as our analysis of the contributions to the VDOS, substituting this into Eq. \eqref{eqn:integral} allows us to identify six contributions to the thermal conductivity coming from each current auto- and cross-correlation function between types of atoms:

\begin{equation}
  \kappa_L=\kappa_{\mathrm{Zn\left(1\right)}}+\kappa_{\mathrm{Zn\left(2\right)}}+\kappa_{\mathrm{Sb}}+
  \kappa_{\substack{\mathrm{Zn\left(1\right)}/\\\mathrm{Zn\left(2\right)}}}+
  \kappa_{\substack{\mathrm{Zn\left(1\right)}/\\\mathrm{Sb}}}+
  \kappa_{\substack{\mathrm{Zn\left(2\right)}/\\\mathrm{Sb}}},
  \label{eqn:contributions}
\end{equation}

\noindent with
\begin{equation}
\kappa_{a/b} \propto \sum\limits_{m=1}^M\frac{1}{L-m}\sum\limits_{n=1}^{L-m} J_a^{\left(z\right)}\left(n\Delta t\right) J_b^{\left(z\right)}\left(0\right).
  \label{eqn:integral}
\end{equation}

\begin{figure}
\centerline{\includegraphics[width=13cm]{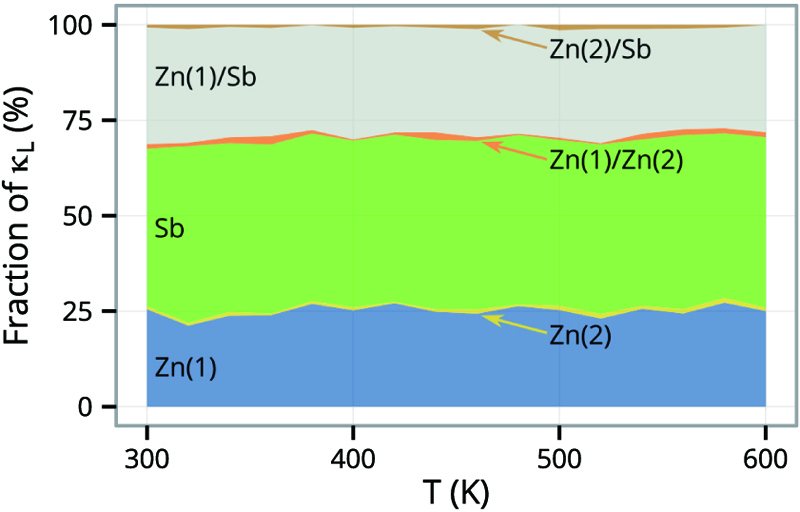}}
\caption{\label{fig3} Percentual contribution of each pair of atom types to the total lattice thermal conductivity. Each contribution is proportional to the corresponding heat current auto- or cross-correlation function.}
\end{figure}

A completely analogous decomposition is often performed for ionic electrical conduction,\cite{mendez-morales_md_2013} where neglecting cross-terms leads to the well-known Nernst-Einstein relation. \cite{hansentheory2006} However, whereas that approximation can be justified for ionic conduction, in the present context the situation is quite different, as illustrated in Figure \ref{fig3}. In fact, at any given temperature the dominating contributions to $\kappa_L$ come not only from the thermal current auto-correlation functions of Zn(1) and Sb, but also from the Zn(1)/Sb cross-term. Since from analysis of the VDOS it was concluded that Zn(1)/Sb coupling is hindered by the fact that they oscillate with frequencies in very different ranges, this third contribution can still be expected to be small in absolute terms. Hence, so can the other two, comparable to it. A low lattice thermal conductivity results, as detected in both calculation and experiment.

In contrast, the three remaining terms, all involving Zn(2), are almost insignificant. The picture that emerges is completely consistent with the discussion above: interstitial atoms do not contribute to thermal conduction and in fact hinder it. This pattern, contrary to the usual behavior of crystalline atoms, sets \ce{Zn4Sb3} apart from more usual materials.

In summary, our results on the thermal conductivity and the projected vibrational densities of states suggest that: 1. Loose coupling between Zn and Sb atoms leads to inefficient thermal conduction; 2. Rattling or diffusive motion of interstitial Zn atoms lowers thermal conductivity even more. These two points have been double-checked by an estimation of thermal conductivity decomposed into contributions from each thermal current auto- and cross-correlation function between different types of atoms. These simulated features clearly point to a significant phonon anharmonicity in $\beta-$\ce{Zn4Sb3}, and provide a clue to the physical origin of the extremely low thermal conductivity of \ce{Zn4Sb3} that approaches the amorphous limit. Other crystalline compounds with loosely coupled atoms exhibiting diffusive behavior can likewise be expected to feature low thermal conductivities, a finding that could suggest new lines of research for experimentalists.

\begin{acknowledgments}
This work was supported by the National Basic Research Program of China (Grant No. 2012CB619402) and the National Natural Science Foundation of China (Grant No. 11204228). We thank Prof. Ju Li at MIT for helpful discussion.
\end{acknowledgments}

\end{document}